\documentclass[showpacs,preprintnumbers,amsmath,amssymb,twocolumn,aps,pre]{revtex4-1}
\usepackage{default}
\usepackage{amssymb}
\usepackage{color}
\usepackage{amsmath}
\usepackage{mathrsfs}
\usepackage{graphicx}
\usepackage{subfig}
\usepackage{epsfig}

\begin{document}
\title{Bifurcation structure of periodic patterns in the Lugiato-Lefever equation with anomalous dispersion}
\author{P. Parra-Rivas$^{1,2,3}$, D. Gomila$^{3}$,  L. Gelens$^{1,2}$ and E. Knobloch$^4$}

\affiliation{ 
 $^1$Laboratory of Dynamics in Biological Systems, KU Leuven Department of Cellular and Molecular Medicine, University of Leuven, B-3000 Leuven, Belgium\\
$^2$Applied Physics Research Group, APHY, Vrije Universiteit Brussel, 1050 Brussels, Belgium\\
  $^{3}$Instituto de F\'{\i}sica Interdisciplinar y Sistemas Complejos, IFISC (CSIC-UIB), Campus Universitat de les Illes
  Balears, E-07122 Palma de Mallorca, Spain\\
  $^4$Department of Physics, University of California, Berkeley CA 94720, USA}

\date{\today}


\begin{abstract}
We study the stability and bifurcation structure of spatially extended patterns arising in nonlinear optical resonators with a Kerr-type nonlinearity and anomalous group velocity dispersion, as described by the Lugiato-Lefever equation. While there exists a one-parameter family of patterns with different wavelengths, we focus our attention on the pattern with critical wave number $k_c$ arising from the modulational instability of the homogeneous state. We find that the branch of solutions associated with this pattern connects to a branch of patterns with wave number $2k_c$. This next branch also connects to a branch of patterns with double wave number, this time $4k_c$, and this process repeats through a series of 2:1 spatial resonances. For values of the detuning parameter approaching $\theta=2$ from below the critical wave number $k_c$ approaches zero and this bifurcation structure is related to the foliated snaking bifurcation structure organizing spatially localized bright solitons. Secondary bifurcations that these patterns undergo and the resulting temporal dynamics are also studied.
\end{abstract}
\maketitle

\section{Introduction}
Since the formulation in 1987 of the Lugiato-Lefever (LL) model describing light propagation in nonlinear optical Kerr cavities \cite{lugiato_spatial_1987}, the existence and origin of spatially extended patterned solutions has been widely studied in both temporal and spatial systems \cite{Firth_patterns1,Firth_patterns2,Scroggie_Tlidi,Tlidi_96, Haelterman,Gomila_hexa1}. In the LL model, it was shown that patterns arise through a Turing instability, usually referred to as a modulational instability (MI) in the optics context \cite{Kapral,Turing,Castets,Cross}. In this type of instability a homogeneous steady state (HSS) becomes unstable to perturbations with a given wavelength, which then further develops into an ordered modulated structure: a {\it pattern}.

In recent years, dissipative structures arising in the one-dimensional LL model have been studied extensively because of their intimate connection to frequency combs in microresonators driven by a continuous wave laser \cite{Haelterman,coen_modeling_2013,chembo_spatiotemporal_2013}. Such frequency combs correspond to the frequency spectrum of localized or extended light patterns that circulate inside the cavity \cite{leo_nature,herr_universal_2012,Leo_OE_2013,Parra-Rivas_PRA_KFCs,godey_stability_2014}, and can be used for a wide variety of applications \cite{kippenberg_microresonator}. In this work, we study the stability and bifurcation structure of extended patterns in the LL model,
\begin{equation}\label{LLE}
 \partial_tA=-(1+i\theta)A+i\nu\partial_x^2A+i|A|^2A+\rho,
\end{equation}
where $\rho$ and $\theta$ are real control parameters representing normalized energy injection and frequency detuning, respectively. We focus here on the anomalous group velocity dispersion (GVD) regime and therefore set $\nu=1$ throughout this work. We study patterns with the critical wave number $k_c$ introduced below, originating from the modulational instability. For the parameter values for which the patterns are subcritical, this bifurcation also leads to the formation of localized structures. For a detailed study of the bifurcation structure of such localized states in the LL model, we refer to \cite{Parra_Rivas_P1}.

This paper is organized as follows. In Section~\ref{sec:1}, we perform the linear stability analysis of the HSS solution with respect to spatially periodic perturbations. This not only reveals the modulational instability, but more generally indicates which perturbation wave numbers lead to instabilities and pattern formation. Next, in Section~\ref{sec:2}, we show how analytical expressions for weakly nonlinear pattern solutions can be found near certain bifurcations. Later, in Section~\ref{sec:3}, we numerically track those analytical solutions to values of the pump parameter $\rho$ away from those bifurcation points, thus revealing the bifurcation structure of the patterns for a fixed value of the detuning. In Section~\ref{sec:4} we study how this bifurcation structure changes as the parameter space defined by the cavity detuning $\theta$ and the pump $\rho$ is traversed, and present phase diagrams showing parameter regimes with distinct pattern behavior. In Section ~\ref{sec:5} a linear stability analysis of the pattern solutions is performed, and the different secondary instabilities that these states undergo are discussed. Finally, in Section~\ref{sec:6} we give some concluding remarks.

\section{Linear stability analysis of the homogeneous steady states}
\label{sec:1}

The HSS solutions $A_0$ can be found by solving the classic cubic equation of dispersive optical bistability, namely
\begin{equation}\label{HSS}
 I_0^3-2\theta I_0^2+(1+\theta^2)I_0=\rho^2,
\end{equation}
where $I_0\equiv|A_0|^2$. The solutions in real variables ($U_0 =$ Re$[A_0], V_0 =$ Im$[A_0]$) are given by
\begin{equation}\label{hom_real}  
\left[\begin{array}{c}
U_0 \\ V_0\end{array}\right]=\left[\begin{array}{c}
\displaystyle\frac{\rho}{1+(I_0-\theta)^2} \\ \displaystyle\frac{(I_0-\theta)\rho}{1+(I_0-\theta)^2}\end{array}\right].
\end{equation}
For $\theta<\sqrt{3}$, Eq.~(\ref{HSS}) is single-valued and hence the system is monostable. In contrast,
for $\theta>\sqrt{3}$, Eq.~(\ref{HSS}) is triple-valued. The transition between the three different solutions occurs via a pair of saddle-node bifurcations SN$_{b}$ and SN$_{t}$ located at 
\begin{equation}
 I_{t,b}\equiv|A_{t,b}|^2=\frac{2\theta}{3}\pm\frac{1}{3}\sqrt{\theta^2-3},
\end{equation}
and these arise from a cusp or hysteresis bifurcation at $\theta=\sqrt{3}$. In what follows, we denote the bottom solution branch (from $I_0=0$ to $I_b$) by $A_0^b$, the middle branch between $I_b$ and $I_t$ by $A_0^m$, and the top branch by $A_0^t$ ($I_0>I_t$). 

A linear stability analysis of the HSS solution with respect to spatially periodic perturbations of the form
\begin{equation}
\left[\begin{array}{c}
U\\V
\end{array}\right]=\left[\begin{array}{c}
U_0\\V_0
\end{array}\right]+\epsilon\left[\begin{array}{c}
u_1(x,t)\\v_1(x,t)
\end{array}\right]+\mathcal{O}(\epsilon^2),
\end{equation}
where $|\epsilon|\ll1$ and
\begin{equation}\label{perturbations}
\left[\begin{array}{c}
u_1\\v_1
\end{array}\right] =\left[\begin{array}{c}
a_k\\b_k
\end{array}\right]e^{ikx+\Omega t}+c.c.,
\end{equation}
leeds to the dispersion relation
\begin{equation}
 \Omega(k)=-1\pm\sqrt{4I_0\theta-3I_0^2-\theta^2+(4I_0-2\theta) k^2- k^4}. 
\end{equation}
Here $\Omega(k)$ is the linear growth rate of a perturbation with wave number $k$.

In the linear approximation, the superposition principle applies and therefore any pattern solution of the problem can be written as the linear combination
\begin{equation}
 \left[\begin{array}{c}
u_1\\v_1
\end{array}\right]_{(x,t)}=\displaystyle\sum_{k}\left[\begin{array}{c}
a_k\\b_k
\end{array}\right]e^{ikx+\Omega t}+c.c.,
\end{equation}
where the mode amplitudes $a_k$, $b_k$ depend on the parameters $\theta$ and $\rho$.
The growth $\Omega(k)$ will in general be positive for wave numbers within an interval $[k^-,k^+]$, where the wave numbers $k^-$ and $k^+$ depend on $I_0$ and solve the quadratic equation
\begin{equation}\label{conditon1}
 k^4-(4I_0-2\theta)k^2+3I_0^2+\theta^2-4I_0\theta-1=0.
\end{equation}
Any mode within this interval will grow, and the profile of the pattern arising from random noise will be dominated by the most unstable mode $k_u$ defined by the condition $\Omega'(k_u)\equiv\frac{d\Omega}{dk}\large|_{k_u}=0$, giving
\begin{equation}\label{condition2}
k_u=\sqrt{2I_0-\theta}.
\end{equation}

The loss of stability occurs at a critical value of $k_c$ where the growth rate first reaches zero, i.e., when conditions (\ref{conditon1}) and (\ref{condition2}) are satisfied simultaneously. This transition is called a Turing \cite{Kapral,Turing,Castets,Cross} or modulational instability (MI), and occurs at $I_0=I_c$, $k=k_c$, where
\begin{equation}
I_c=1,\qquad k_c=\sqrt{2-\theta}.
\end{equation}
Evidently, this transition is only found when $\theta<2$. The condition $I_0=I_c$ defines a line in the parameter space $(\theta,\rho)$ given by 
\begin{equation}
\rho_c=\sqrt{1+(1-\theta)^2}.
\end{equation}

\begin{figure}
\centering
\includegraphics[scale=1]{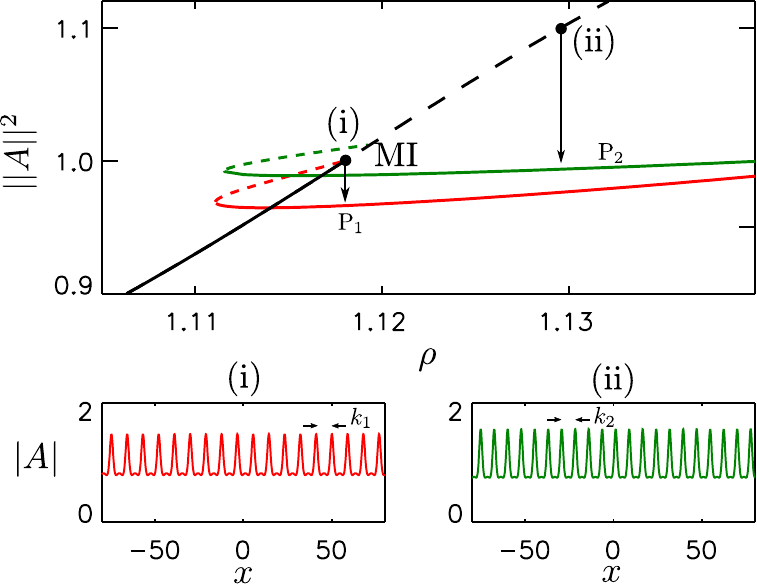}
\caption{(Color online) The stable HSS (black solid line) is destabilized at the modulational instability MI. Close to MI (i), the unstable HSS evolves to the pattern branch P$_1$ (red) consisting of stationary patterns with wave number $k_1 =  8.889 \approx k_c = 8.886$. Further away from MI (ii), the unstable HSS evolves into a different pattern branch P$_2$ (green), now characterized by patterns with wave number $k_2 =  7.620 \approx k_u = 7.510$. Stable (unstable) solutions are denoted by solid (dashed) lines. Here $\theta = 1.5$ and $L=160$.}
\label{MI_patterns_intro}
\end{figure}

Figure~\ref{MI_patterns_intro} illustrates how the HSS destabilizes when the pump parameter $\rho$ exceeds $\rho=\rho_c$ and how the pattern state is subsequently reached. The wave number of this pattern changes with the pump parameter as does the most unstable wave number [see Eq.\ (\ref{condition2})]. Close to the MI the HSS develops into a pattern that lies on a branch of pattern solutions with wave number close to $k_c$, originating near MI. For larger values of the pump, however, the selected pattern belongs to a pattern branch corresponding to a wave number close to the fastest growing wave number $k_u$. This observation highlights the fact that the pattern branches form a continuum, parametrized by the wavenumber $k\in[k^-,k^+]$, with the wave number selected by nonlinear processes that depend on the system parameters. In this work we restrict attention to pattern branches corresponding to the critical wave number $k_c$ and its harmonics, and describe their bifurcation structure in some detail. The study of patterns with other wave numbers is left for future work.

\begin{figure}
\centering
\includegraphics[scale=1]{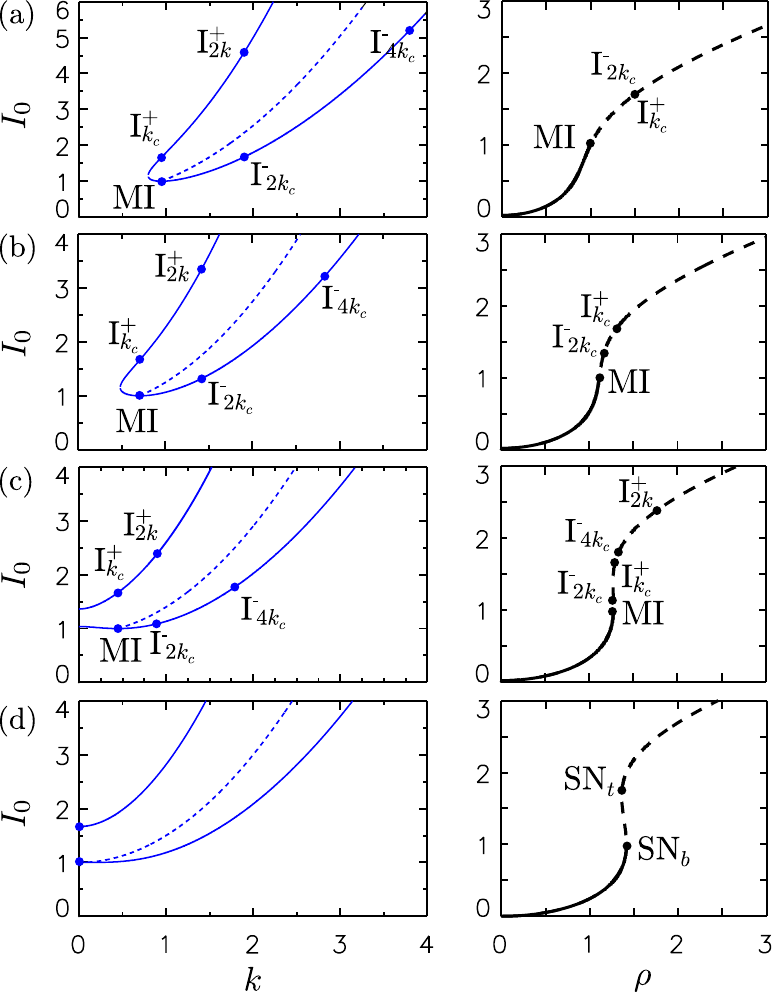}
\caption{(Color online) Left: Marginal instability curves for (a) $\theta=1.1$, (b) $\theta=1.5$, (c) $\theta=1.8$ and (d) $\theta=2.0$ (d). Right: The HSS solutions corresponding to the same values of $\theta$. Solid (dashed) lines represent stable (unstable) HSSs with respect to perturbations of the form (\ref{perturbations}). The locations I$^{\pm}_{k}$ corresponding to instabilities with wave number $k$ are indicated using solid circles. The dashed line inside the marginal instability curves in the left panels represents the most unstable mode $k=k_u$.}
\label{marginal1}
\end{figure}

Before turning to the bifurcation structure of pattern solutions, we start our analysis by studying the set of points $k^-$ and $k^+$ satisfying Eq.~(\ref{conditon1}). These points define the so-called {\it marginal stability curve} defined by
\begin{equation}\label{marginal}
{\rm I}^{\pm}_k(\theta)=\frac{2}{3}(\theta+k^2)\pm\frac{1}{3}\sqrt{\theta^2+k^4+2\theta k^2-3}.
\end{equation}
The marginal stability curves are shown in the panels on the left of Fig.~\ref{marginal1} for increasing values of the detuning $\theta$. The HSS solutions at the corresponding values of $\theta$ are shown in the panels on the right, with solid (dashed) lines representing the HSS solutions that are stable (unstable) against perturbations of the form (\ref{perturbations}). For a fixed value of $\theta$, and for a given wave number $k'$, the HSS solution is unstable if ${\rm I}^{-}_{k'}(\theta)<{\rm I}_0< {\rm I}^{+}_{k'}(\theta)$ and stable otherwise. Thus, for a given wave number $k=k_c$ a pattern P$_{k_c}$ bifurcates from the points I$^{\pm}_{k_c}(\theta)$ indicated in Fig.~\ref{marginal1} and similarly for patterns with wavenumber $2k_c$, $4k_c$ etc.

\begin{figure}
\centering
\includegraphics[scale=0.97]{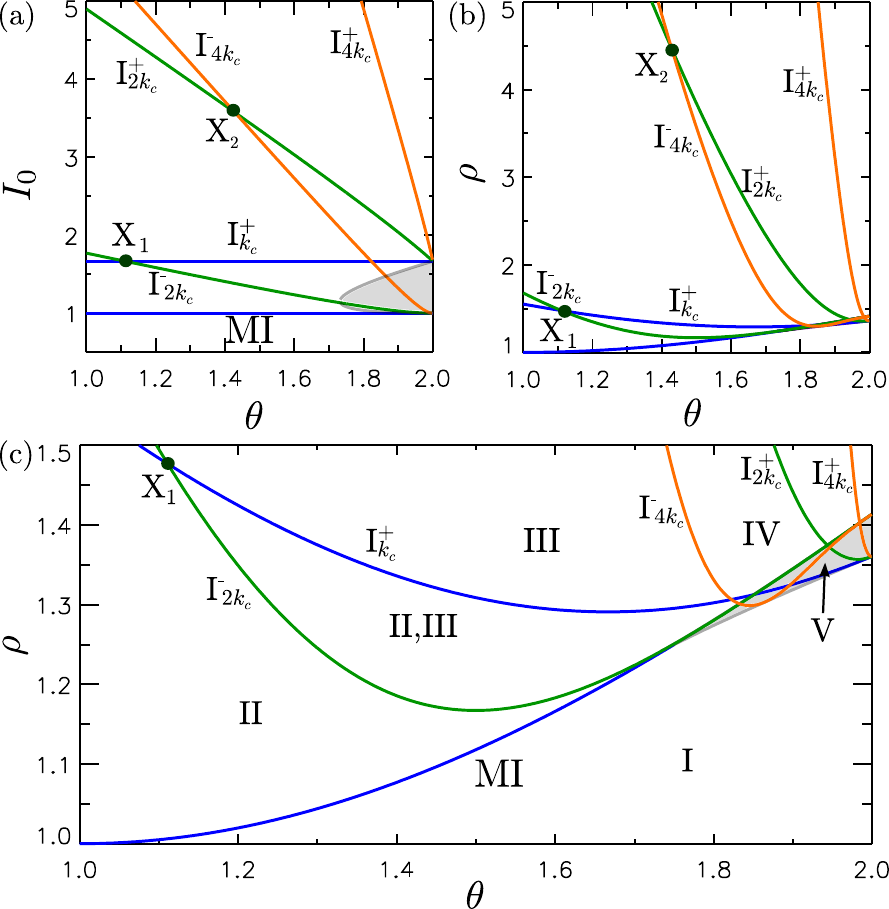}
\caption{(Color online) (a) The instability lines I$^{\pm}_{k_c}$ and the location of the saddle-node bifurcations of the HSSs in the parameter space $(\theta,I_0)$.
(b) Same as (a) but in the parameter space $(\theta,\rho)$. (c) Zoom of (b) showing the main regions with distinct bifurcation behavior (see text).
The labels X$_1$ and X$_2$ indicate codimension-two points. In both (a) and (c) the gray area represents region IV where the system has a bistable response in the HSS solutions.}
\label{parameterb}
\end{figure}

In Fig.~\ref{marginal1}(a), for $\theta=1.1$, the HSS is always stable against perturbations with $k=0$. Furthermore, a pattern with wavenumber $k_c$ bifurcates from the MI at I$^-_{k_c}={\rm I}_c$ and then reconnects with HSS again at I$^+_{k_c}>{\rm I}^-_{k_c}$. Similarly, a pattern with $2k_c$ arises initially from I$^-_{2k_c}$ and reconnects to HSS at I$^+_{2k_c}$. The situation for all subsequent harmonics is similar. As the detuning $\theta$ increases, the different instability points for modes with $k=k_c$ and its harmonics approach each other as the whole tongue of unstable modes shifts to lower values of $k$ [see Fig.~\ref{marginal1}(b)]. This behavior can also be seen in Fig.~\ref{parameterb} where we plot the instability boundaries in the parameter space $(\theta,I_0)$ and $(\theta, \rho)$, respectively, together with the location of the saddle-node bifurcations SN$_b$ and SN$_t$ of the HSS solution. For $\theta<\sqrt{3}$, $A_0$ is always stable against spatially uniform perturbations with $k=0$. In contrast, when $\sqrt{3}<\theta<2$, the response of the HSSs as a function of the pump parameter $\rho$ becomes bistable. In this case, the bottom $A_0^b$ and top $A^t_0$ branches are stable with respect to $k=0$ perturbations, while the middle branch $A_0^m$ is unstable to such perturbations. However, $A^t_0$ and $A^m_0$ are always unstable with respect to $k>0$ perturbations, while $A^b_0$ is only destabilized above $I_0=I_c$. This situation is depicted in Fig.~\ref{marginal1}(c) for $\theta=1.8$, where the tongue of unstable wavenumbers now starts at $k=0$. 

Finally, when the detuning incfeases to $\theta = 2$ from below the instability points I$^{\pm}_{nk_c}$, $n=1,2,\dots$, approach one another until they all collapse at $k=0$ and the MI disappears [see Fig.~\ref{marginal1}(d)]. A similar collapse can be seen in Fig.~\ref{parameterb}, where I$^+_{k_c}$ and I$^-_{2k_c}$, and I$^+_{2k_c}$ and I$^-_{4k_c}$ collide pairwise at the codimension-two bifurcation X$_1$ and X$_2$ located at $(\theta_{{\rm X}_1},\rho_{{\rm X}_1})=(1.1111,1.4768)$, and $(\theta_{{\rm X}_2},\rho_{{\rm X}_2})=(1.4286,4.468)$, respectively. The results presented in Fig.~\ref{marginal1} and Fig.~\ref{parameterb} are limited to $\theta<2$ for which the MI exists and takes place at $I_0=I_c$. When approaching $\theta=2$ from below, the critical wave number approaches zero ($k_c\rightarrow0$), implying that the wavelength of the nascent pattern diverges. Since a pattern with infinite wavelength corresponds to a single peak in the domain, the distinction between patterns and localized structures becomes blurred in this limit. A detailed analysis of how the bifurcation structure of such localized structures changes as one approaches this critical point $\theta=2$ can be found in Ref.~\cite{Parra_Rivas_P1}.

%

At this point we can already identify several distinct solution regimes based on the existence of patterns and the stability of $A_0$:
\begin{itemize}
 \item Region I: The HSS solution $A_0$ is stable. This region spans the parameter space $\rho<\rho_{c}$.
 \item Region II: The pattern P$_{k_c}$ exists between MI and I$^+_{k_c}$, and $A_0$ unstable.
 \item Region III: The pattern P$_{2k_c}$ exists between I$^-_{2k_c}$ and I$^+_{2k_c}$, and $A_0$ is unstable. 
 \item Region IV: The pattern P$_{4k_c}$ exists between I$^-_{4k_c}$ and I$^+_{4k_c}$, and $A_0$ is unstable. 
  \item Region V: Multistability of the HSS $A_0$. $A_0^b$ is stable, while $A_0^t$ and $A_0^m$ are unstable. This region spans the parameter region between SN$_b$ and SN$_t$. The patterns P$_{k_c}$ and P$_{2k_c}$ also exist in this region since they appear subcritically.
\end{itemize}

In the following sections we study how the different patterns reconnect as parameters are varied, and identify the different instabilities these patterns undergo.

\section{Weakly nonlinear pattern solutions}\label{sec:2}

Weakly nonlinear patterns are present in the vicinity of the MI bifurcation at $I_0=I_c$ and can be computed using multiscale perturbation analysis. At leading order in the expansion parameter $\epsilon$, defined by the relation $\rho=\rho_c+\epsilon^2\mu$, the pattern solution is given by  
\begin{equation}\label{pattern_asymp}
\left[\begin{array}{c}
        U\\V
       \end{array}
\right]= \left[\begin{array}{c}
        U_c\\V_c
       \end{array}
\right]+\epsilon\left[\begin{array}{c}
        u_1\\v_1
       \end{array}
\right]+\epsilon^2\left[\begin{array}{c}
        U_2\\V_2
       \end{array}
\right],
\end{equation}
where $U_c$ and $V_c$ correspond to the HSS solution (\ref{hom_real}) at $\rho=\rho_c$, $U_2$ and $V_2$ represent the leading order correction to this HSS, given by
\begin{equation}
 \left[\begin{array}{c}
        U_2\\V_2
       \end{array}
\right]=\frac{\mu}{{\left(\theta^{2} - 2 \, \theta +
2\right)} {\left(\theta - 2\right)}}\left[\begin{array}{c}
\theta^{2} \\-\theta^{2} - \theta + 2\end{array}\right],
\end{equation}
and the space-dependent correction is given by
\begin{equation}
 \left[\begin{array}{c}
        u_1\\v_1
       \end{array}
\right]=2 \left[\begin{array}{c}
        a\\1
       \end{array}
\right]B {\rm cos}(k_c x+\varphi),
\end{equation}
where $\varphi$ is an arbitrary phase, and
\begin{equation}
 a=\frac{\theta}{2-\theta}.
\end{equation}
The amplitude $B$ for the pattern state corresponds to the constant solution of the amplitude equation
\begin{equation}
C_1 B_{XX}+\mu C_2B+C_3B^3=0,
\end{equation}
i.e., 
\begin{equation}
 B=\sqrt{-\delta C_2/ C_3}.
\end{equation}
Here
\begin{equation}
C_1=
-\frac{2 \, {\left(\theta^{2} - 2 \, \theta + 2\right)}}{\theta - 2},
\end{equation}

\begin{equation}       	
C_2=
\frac{2 \, {\left(\theta^{2} - 2 \, \theta +
2\right)}^{\frac{3}{2}}}{{\left(\theta - 2\right)}^{4}},
\end{equation}

\begin{equation}      	
C_3=\frac{4 \, {\left(\theta^{2} - 2 \, \theta +
2\right)}^{2} {\left(30 \, \theta - 41\right)}}{9 \, {\left(\theta -
2\right)}^{6}}.
\end{equation}
It follows that the pattern is supercritical for $\theta<41/30$ but subcritical for $\theta<41/30$, as already predicted in Refs.~\cite{lugiato_spatial_1987,Perinet_Eckhaus}. In the following we refer to this pattern as P$_{k_c}$. Details of the above calculation can be found in Ref.~\cite{Parra_Rivas_P1}.
\begin{figure*}[!t]
\centering
\includegraphics[scale=1]{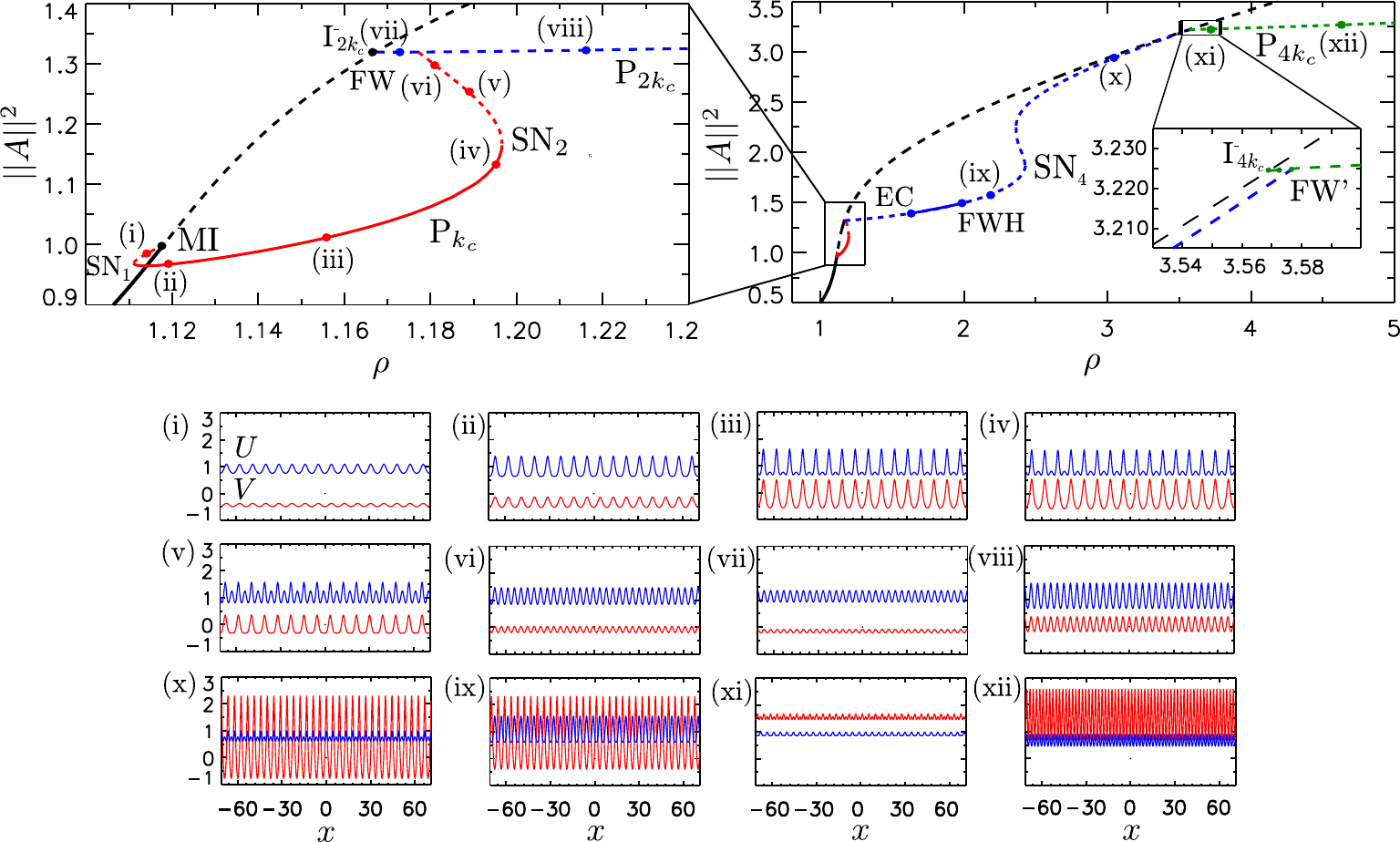}
\caption{(Color online) Bifurcation diagrams for patterns with wave numbers $k_c$, $2k_c$, and 
$4k_c$ for $\theta=1.5$. Solution profiles along the different branches are shown in panels (i)-(xii).}
\label{motiv}
\end{figure*}

\section{Bifurcation structure of patterns}\label{sec:3}

We now present the main features of the bifurcation structure of the pattern states for a fixed value of the detuning, choosing $\theta=1.5$ as a representative value, leaving the study of how this structure is modified as $\theta$ varies to the following section. Starting from the analytical solution (\ref{pattern_asymp}), valid close to the MI bifurcation, we use a numerical continuation algorithm to construct the bifurcation diagram shown in Fig.~\ref{motiv}, showing the intensity $||A||^2$ as a function of the parameter $\rho$. As in Fig.~\ref{MI_patterns_intro}, the black lines represent HSSs, while red, blue and green lines correspond to patterned states with wave number $k_c$, $2k_c$ and $4k_c$, respectively. Furthermore, solid lines denote stable solutions, while dashed lines indicate unstable ones. Different profiles along these branches are shown in panels (i)-(xii). As shown in Fig.~\ref{MI_patterns_intro}, the pattern P$_{k_c}$ with wave number $k_c$ originates at the MI bifurcation. 

While the MI bifurcation corresponds to the point where the HSSs lose stability to temporal perturbations, it is also possible to study this transition in the context of spatial dynamics. Here, the HSS is interpreted as a fixed point in a four-dimensional phase space \cite{Parra_Rivas_P1}, and the MI corresponds to a Hamiltonian-Hopf (HH) bifurcation with eigenvalues $\lambda=\pm ik_c$ of double multiplicity. In this formulation the pattern state corresponds to a periodic orbit, and this orbit bifurcates from HSS at $\rho_c$ (for $\theta<2$) with initial period (wavelength) $2\pi/k_c$. Together with this critical pattern there is a continuous family of patterns with $k\in[k^-,k^+]$ that bifurcates from the HSS solution for $\rho>\rho_c$. Within the spatial dynamics framework the HSS points for $\rho>\rho_c$ are nonhyperbolic and the bifurcations to P$_{2k_c}$, P$_{4k_c}$,$\dots$ have no particular signature within the spatial dynamics point of view. However, linear stability theory in the time domain shows that bifurcations occur whenever the spatial eigenvalues on the imaginary axis are in resonance, $k=nk_c$, where $n$ is an integer. Theory also shows that the primary bifurcation to periodic orbits at $\rho_c$ is accompanied by the simultaneous appearance of a pair of branches of spatially localized structures, provided only that the periodic states bifurcate subcritically. As a result the localized states can be interpreted as portions of the pattern state embedded in a uniform background. The bifurcation structure of such localized structures is studied in detail in Ref.~\cite{Parra_Rivas_P1}.

As the detuning $\theta$ in Fig.~\ref{motiv} is larger than $41/30$, the pattern P$_{k_c}$ is created subcritically and is therefore initially temporally unstable [see profile (i)]. Following this branch away from MI, the pattern grows in amplitude and gains stability at a saddle-node bifurcation SN$_{1}$ [profiles (ii)-(iv)], but loses stability at a second saddle-node SN$_{2}$ [profiles (v)-(vi)]. Once SN$_{2}$ is passed, spatial oscillations (SOs) start to appear in between the peaks in the pattern profile as seen most clearly in profile (v). These SOs correspond to the growth of the second harmonic $2k_c$ of the pattern wave number, and these grow in amplitude with increasing $\rho$ [profile (vi)] until P$_{k_c}$ merges with the pattern P$_{2k_c}$, a state with wave number $2k_c$ (plus harmonics). The merging of these two periodic orbits occurs in a 2:1 spatial resonance \cite{Armbruster,Porter,Proctor}, which in the context of patterns corresponds to a finite wavelength (FW) instability of P$_{2k_c}$ that doubles its wavelength, i.e., to a (spatial) subharmonic instability.

The pattern P$_{2k_c}$ itself bifurcates supercritically from HSS at I$^-_{2k_c}$. Since this branch inherits the unstable eigenvalue of HSS the P$_{2k_c}$ branch is initially unstable. The resulting likewise grows in amplitude as $\rho$ increases [profiles (vii)-(viii)] but at SN$_4$, it folds back and just as for P$_{k_c}$, SOs appear between successive peaks in the profile and the pattern terminates at a FW$'$ point on the P$_{4k_c}$ branch with characteristic wave number $4k_c$ once the amplitude of the SOs reaches that of the original peaks. This new pattern again bifurcates supercritically from the HSS, this time at I$^-_{4k_c}$ [profile (xi)], and is likewise initially unstable before terminating in yet another 2:1 spatial resonance [profile (xii)]. We have identified a whole cascade of such bifurcations involving ever higher harmonics of $k_c$.

Bifurcation theory sheds light on the bifurcation sequence described above. We imagine that the bifurcations to P$_{k_c}$ and P$_{2k_c}$ occur in close succession and so look for solutions in the form $(U,V)\propto z_1\exp ik_cx+z_2\exp 2ik_cx+{\rm c.c.}+{\rm h.o.t}$. The complex amplitudes $z_1$, $z_2$ then satisfy the equations \cite{Armbruster,Proctor,Porter}
\begin{equation}\label{sr_eq}
\begin{array}{l}
\dot{z}_1=\mu z_1+c_1{\bar z}_1z_2+(e_{11}|z_1|^2+e_{12}|z_2|^2)z_1+\dots\\\\
\dot{z}_2=(\mu-\nu) z_2+c_2z_1^2+(e_{21}|z_1|^2+e_{22}|z_2|^2)z_2+\dots
\end{array}
\end{equation}
We see that for fixed $\nu>0$ the HSS solution $(z_1,z_2)=(0, 0)$ loses stability in succession to modes with wave numbers $k_c$, $2k_c$ as $\mu$ increases. We also see that the equations admit a pure P$_{2k_c}$ solution $(0,z_2)$ but that the P$_{k_c}$ state acquires a contribution with wave number $2k_c$ as soon as $\mu>0$, exactly as observed in the figure, i.e., the mode starting out as $(z_1,0)$ is in fact a mixed mode $(z_1,z_2)$ as soon as $\mu>0$. Moreover, as $\mu$ increases the contribution from the amplitude $z_2$ grows and the mixed mode terminates on the $(0,z_2)$ branch of pure wave number $2k_c$ states, also as observed. The latter is a 2:1 resonance since at this bifurcation a pure mode with wave number $2k_c$ bifurcates into a mixed mode with a contribution from wave number $k_c$. We can therefore think of this bifurcation as a subharmonic instability in space.  

In the next section, we explore how the bifurcation structure connecting P$_{k_c}$ with all its harmonics is modified when the cavity detuning $\theta$ varies.

\begin{figure}[!t]
\centering
\includegraphics[scale=1]{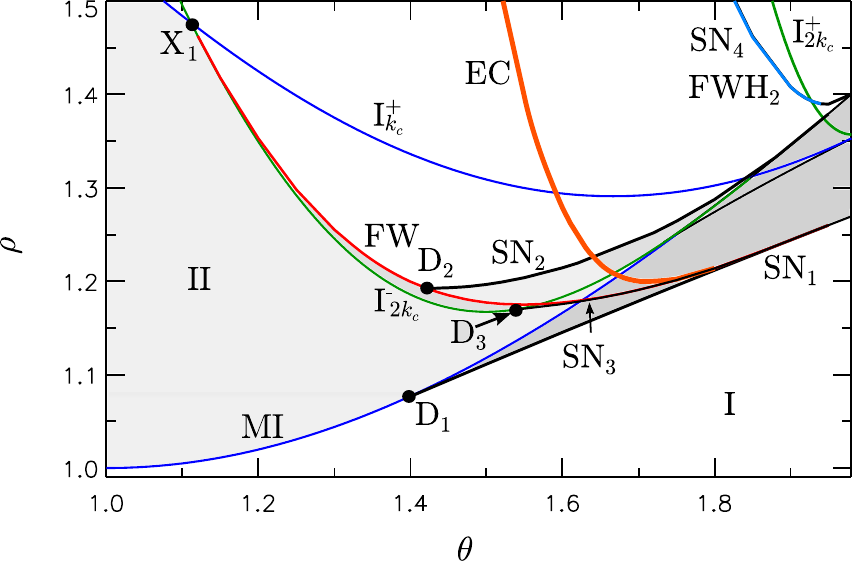}
\caption{(Color online) Phase diagram in the $(\theta,\rho)$ parameter space $(\theta,\rho)$ showing the main bifurcations of HSS and the pattern states. The region of bistability between $A^b_0$ and P$_{k_c}$ is indicated in dark gray, while the wider region of stability of P$_{k_c}$ is colored in light gray. The symbol $\bullet$ represents the codimension-two points X$_1$, D$_1$, D$_2$, and D$_3$.}
\label{parameter}
\end{figure}

\section{Patterns in the $(\theta,\rho)$ plane}\label{sec:4}

\begin{figure}[t!]
\centering
\includegraphics[scale=0.91]{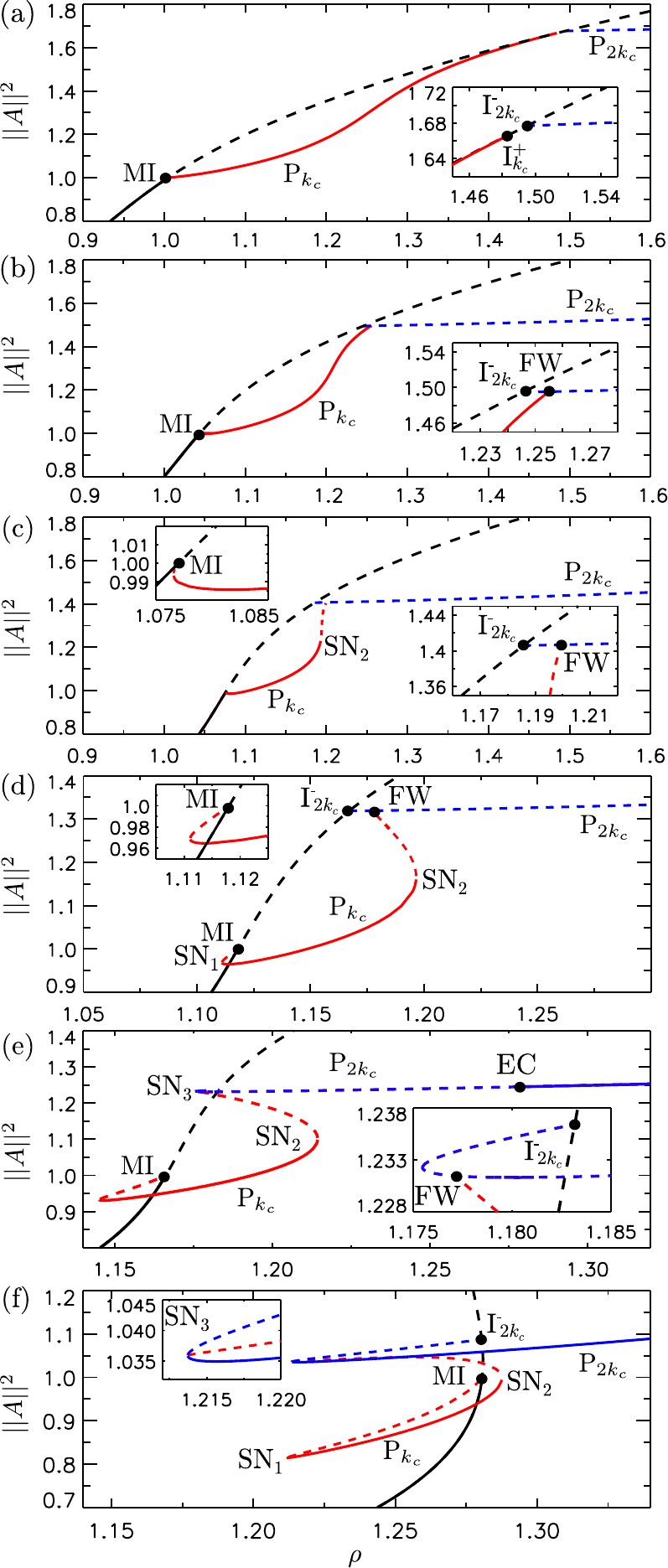}
\caption{(Color online) Bifurcation diagrams corresponding to (a) $\theta=1.1$, (b) $\theta=1.3$, (c) $\theta=1.4$, (d) $\theta=1.5$, (e) $\theta=1.6$ and (f) $\theta=1.8$. Red lines correspond to P$_{k_c}$ and the blue lines to P$_{2k_c}$. Panels (a) and (b) show the situation before and after the codimension-two point X$_1$. Panels (b) and (c) show the transition from supercritical to subcritical bifurcation of pattern P$_{k_c}$ via a degenerate HH at $\theta=41/30$.  For $\theta=1.5$ [panel (d)] P$_{2k_c}$ bifurcates supercritically from HSS at I$^-_{2k_c}$. In contrast, for $\theta=1.6$ [panel (e)] P$_{2k_c}$ emerges subcritically. Solid (dashed) lines indicate stable (unstable) branches.}
\label{bif_dia_several}
\end{figure}

Figure~\ref{parameter} shows the different bifurcation lines and dynamical regions introduced in the previous sections in the $(\theta,\rho)$ parameter space. As this phase diagram is quite dense and therefore difficult to interpret, we show the changes of the bifurcation structure as a function of the pump $\rho$ for increasing values of the detuning $\theta$ in Fig.~\ref{bif_dia_several}. 

For small values of $\theta$ [Fig.~\ref{bif_dia_several}(a), $\theta=1.1<41/30$], the pattern P$_{k_c}$ (red line) bifurcates supercritically from MI at $I_0=I_c$ and connects back to the HSS at I$^+_{k_c}$; P$_{2k_c}$ (blue line) is disconnected from P$_{k_c}$ and bifurcates from I$^-_{2k_c}$ and then extends to higher values of $\rho$ before connecting with HSS at I$^+_{2k_c}$. When $\theta$ increases, I$^+_{k_c}$ and I$^-_{2k_c}$ collide at a codimension-two bifurcation labeled X$_1$, after which the P$_{k_c}$ and P$_{2k_c}$ branches connect to one another with a FW instability originating in X$_1$. This is the 2:1 spatial resonance mentioned in the previous section. This situation is shown in Fig.~\ref{bif_dia_several}(b). Here both patterns emerge supercritically from the HSS state, with P$_{k_c}$ stable and P$_{2k_c}$ initially unstable. However, the latter can change stability through subsequent Eckhaus (EC) and finite-wavelength-Hopf (FWH) instabilities (see Fig.~\ref{parameter}), resulting in more complex scenarios studied in Section~\ref{sec:4}.


At $\theta=41/30$, the bifurcation to P$_{k_c}$ is a degenerate HH bifurcation denoted in Fig.~\ref{bif_dia_several}(c) by D$_1$. For $\theta>41/30$ the bifurcation is subcritical as shown in Fig.~\ref{bif_dia_several}(c) for $\theta=1.4$. Here, P$_{k_c}$ is initially unstable but acquires stability at a saddle-node labeled SN$_1$. This branch then connects with P$_{2k_c}$ at FW. Thus a parameter regime is present in which $A^b_0$ and P$_{k_c}$ coexist stably. As a result localized structures (LS) are also present and these are organized in a so-called homoclinic snaking structure \cite{Parra_Rivas_P1,Coullet,Gomila_Schroggi,Woods1999,Burke_Knobloch}. These LS are found in the region of bistability between $A^b_0$ and P$_{k_c}$ colored in dark gray in Fig.~\ref{parameter}, while the wider region of stability of P$_{k_c}$ is colored in light gray. For $\theta=1.5$ [Fig.~\ref{bif_dia_several}(d)], the situation remains similar, but P$_{k_c}$ now bifurcates subcritically from FW, i.e. an unstable pattern emerges from FW and gains stability at SN$_{2}$. This change in direction of branching is also associated with a codimension-two point, this time labeled D$_2$ (Fig.~\ref{parameter}). As a result, the upper portion of the P$_{k_c}$ branch is stable between SN$_1$ and SN$_2$ while the lower parts between MI and SN$_1$ and between FW and SN$_2$ are both unstable.

For $\theta=1.6$ [Fig.~\ref{bif_dia_several}(e)], the HSS branch is still monotonic but P$_{2k_c}$ now also emerges subcritically, having crossed another degeneracy at D$_3$ (Fig.~\ref{parameter}). This leads to the creation of a saddle-node bifurcation SN$_3$ on the P$_{2k_c}$ branch similar to SN$_1$ on the P$_{k_c}$ branch. At the same time an Eckhaus bifurcation moves in from larger values of $\rho$, stabilizing the large $\rho$ part of the P$_{2k_c}$ branch. With further increase in $\theta$ the EC point collides with FW, and the whole P$_{2k_c}$ branch beyond FW becomes stable. For yet larger $\theta$ the FW point moves towards SN$_3$ so that P$_{k_c}$ now terminates on P$_{2k_c}$ at SN$_3$ and the P$_{2k_c}$ branch stable from SN$_3$ towards larger $\rho$. This multiple bifurcation occurs for $\theta\approx 1.72$ but is not analyzed in this work. Figure~\ref{bif_dia_several}(f) shows the resulting bifurcation diagram when $\theta=1.8$. Since this value of $\theta$ exceeds $\sqrt{3}$ the HSS branch is no longer monotone, with I$^-_{k_c}$ lying below the resulting fold SN$_b$ and I$^-_{2k_c}$ above it.



In Figs.~\ref{parameter} and \ref{bif_dia_several}, we focus on the bifurcations associated with P$_{k_c}$ and P$_{2k_c}$, although very similar transitions occur between P$_{2k_c}$ and P$_{4k_c}$, P$_{4k_c}$ and P$_{8k_c}$, and so on. This scenario resembles foliated snaking of localized structures that appears for $\theta > 2$ \cite{Parra_Rivas_P1}. Since $k_c \rightarrow 0$ as $\theta \rightarrow 2$ from below, in a finite system a pattern with domain-size wavelength becomes indistinguishable from a single peak localized structure present for $\theta>2$, i.e., in the limit $\theta \rightarrow 2$ P$_{k_c}$ becomes a single peak LS, P$_{2k_c}$ becomes a two peak LS, etc. thereby reproducing precisely the foliated snaking bifurcation scenario.

A similar pattern organization exists for patterns with wave number $k\ne k_c$, implying that the complete scenario is fundamentally complex. A detailed study of secondary bifurcations of patterns with wave numbers $k\ne k_c$ is therefore left for future work.

\section{Linear stability analysis of the pattern solutions}\label{sec:5}

The preceding section has highlighted the importance of a secondary wavelength changing instability called the Eckhaus instability. This is a long wavelength instability, with domain-size wavelength, and its nonlinear evolution generally leads to the generation of a phase slip whereby a new roll is injected (or annihilated) at the location of the phase slip, followed by relaxation of the new pattern towards a periodic structure with a new and different wavelength in the domain \cite{KramerZimmermann,BarkleyTuckerman}.

The traditional approach to describing the Eckhaus is based on the use of an amplitude equation, the Ginzburg-Landau equation, that describes the pattern-forming instability close to the primary pattern-forming bifurcation, assumed to be supercritical \cite{BarkleyTuckerman,Eckhaus_saliya}. As a result the predictions concerning the onset and evolution of the Eckhaus instability are valid only when the instability sets in close to the primary instability. We have seen that in the present case this is not so -- in some cases the primary bifurcation is subcritical and the analysis of the Eckhaus instability is then substantially modified \cite{KaoKnobloch}. 
For this reason we apply here a technique described in \cite{Harkness1,Gomila_hexa1} that permits us to compute the onset of the Eckhaus instability for finite amplitude fully nonlinear spatially periodic patterns. The technique is necessarily numerical but allows us to find and characterize, as a function of $\theta$, $\rho$, and $k$, the secondary bifurcations introduced in Section~\ref{sec:4}. Similar numerical studies have been performed in the context of fluid mechanics in Ref.~\cite{Mercader} and for supercritical patterns within the LL equation in Ref.~\cite{Perinet_Eckhaus}.

The stationary patterns, hereafter $A_p=(U_p, V_p)$, can be written as a Fourier modal expansion
\begin{equation}\label{sta_ansatz2}
A_p(x)=\displaystyle\sum_{m=0}^{N-1} a_m e^{imkx},
\end{equation}
with $k$ the wave number of the pattern, $a_m$ the complex amplitude of the Fourier mode with wave number $mk$, and $N$ the number of Fourier modes retained in the analysis. To study the linear stability of such a pattern state, one must first linearize Eq.~(\ref{LLE}) around the state (\ref{sta_ansatz2}). Writing $A(x,t)=A_p(x)+\epsilon \delta A(x,t)$, $\epsilon\ll1$, leads to the following leading order equation for the perturbation $\delta A$:
\begin{equation}\label{linear_complex}
 \partial_t\delta A=-(1+i\theta)\delta A+i\partial_x^2\delta A+2i|A_p|^2\delta A+iA_p^2\delta A^*.
\end{equation}

Owing to the periodicity of $A_p$, we can apply the Bloch ansatz and write the eigenmodes of this equations as Bloch waves
\begin{equation}\label{sta_ansatz1}
\delta A(x,t)=e^{iqx}\delta a(x,t,q)+e^{-iqx}\delta a^*(x,t,-q),
\end{equation}
where $\delta a$ has the same spatial period as the pattern $A_p$ and can be written in the form
\begin{equation}
 \delta a(x,t,q)=\displaystyle\sum_{m=0}^{N-1} \delta a_m(t,q)e^{ikmx}.
\end{equation}
Inserting Eqs.~(\ref{sta_ansatz2}) and (\ref{sta_ansatz1}) in Eq.~(\ref{linear_complex}) leads to a set of linear equations for the complex amplitudes $\delta a_n^{\pm}\equiv\delta a_n(t,\pm q)$, namely
\begin{multline}
\frac{d}{dt} \delta a^{\pm}_n=-(1+i\theta)\delta a^{\pm}_n-i(kn\pm q)^2\delta a^{\pm}_n+\\
i\sum_{l,m=0}^{N-1}a_la_m^*\delta a^{\pm}_{n-l+m}+i\sum_{l,m=0}^{N-1}a_la_m\delta a^{* \pm}_{-n+l+m}.
\end{multline}
This equation has the form
\begin{equation}
 \partial_t{\Sigma}_n(t,q)=L(a_n,q)\Sigma_n(t,q),
\end{equation}
where
$$\Sigma_n(t,q)\equiv(\delta a^+_0,\cdots,\delta a^+_{N-1},\delta a^{*-}_0,\cdots,\delta a^{*-}_{N-1}).$$ Thus, the linear stability analysis of $A_p(x)$ reduces to finding the 2N eigenvalues $\lambda_n(q)$ of the $N\times N$ matrix $L(a_n,q)$ and the corresponding eigenvectors, for each value of $q$. For more details, see Refs.~\cite{Gomila_hexa1,Harkness1,Harkness2}. The eigenvalues for a given $q$ determine the stability of the pattern against perturbations containing wave numbers $k\pm q$ for any $k$. For this purpose it is sufficient to consider only $q$ values inside the first Brillouin zone. Any perturbation with wave number $q'$ outside the Brillouin zone is equivalent to another with $q=q'+k$. In solid state physics this representation is described as the {\it reduced zone scheme} \cite{Ashcroft_Mermin}.

Using this technique we characterize how the eigenspectrum of $L(A_p)$ changes as a function of $q$ for different values of $(\theta,\rho)$, and predict the different secondary bifurcations that a pattern with wave number $k$ undergoes.

\begin{figure}[!t]
\centering
\includegraphics[scale=1]{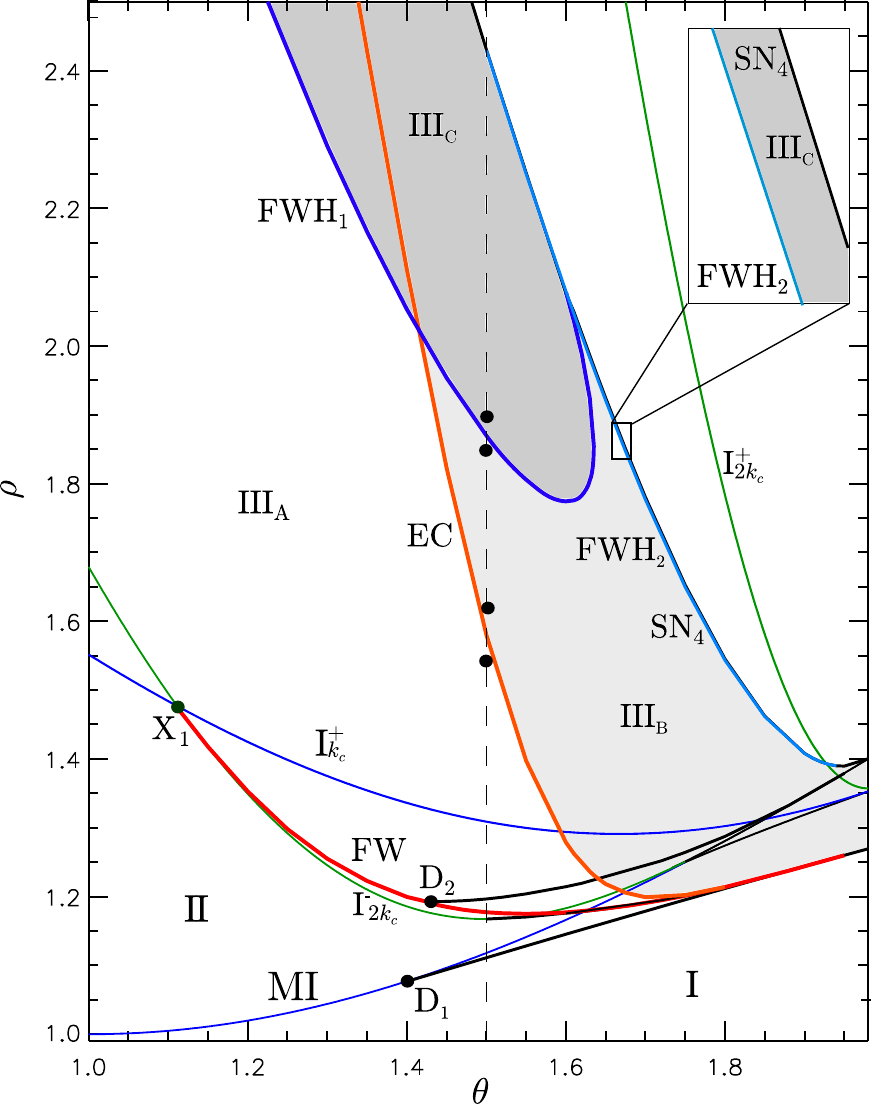}
\caption{(Color online) (a) Phase diagram in $(\theta,\rho)$ parameter space showing an enlargement of the diagrams shown in Fig.~\ref{parameter} focusing on the main stability regions of P$_{2k_c}$ labeled III$_{\rm A,...,C}$. The dashed line at $\theta=1.5$ refers to the slice of this diagram shown in Fig.~\ref{diagrama_EC}. On top of these lines, the symbol $\bullet$ corresponds to points where the stability analysis of the patterns shown in Section~\ref{sec:4} was performed. }
\label{parameter2}
\end{figure}

Figure~\ref{parameter2} shows an enlarged version of the phase diagram in Fig.~\ref{parameter}. We see that the pattern P$_{k_c}$ is stable everywhere between SN$_1$ and SN$_2$. However, P$_{2k_c}$ undergoes three types of secondary instability indicated in Figs.~\ref{parameter} and \ref{parameter2} by the lines EC (Eckhaus), FW (finite-wavelength), and FWH (finite-wavelength-Hopf). These bifurcations divide region III [see Fig.~\ref{parameter2}] into the following subregions:
\begin{itemize}
 \item Region III$_{\rm A}$: The pattern P$_{2k_c}$ is Eckhaus unstable. 
 This region spans the parameter space between I$_{2k_c}^-$ and SN$_3$ from below, and FWH$_1$ and EC from above.
 \item Region III$_{\rm B}$: P$_{2k_c}$ is stable between EC and SN$_3$ from below, and FWH$_1$ and FWH$_2$ from above.
 \item Region III$_{\rm C}$: P$_{2k_c}$ oscillates in time and in space. This region spans the parameter space inside the region defined by FWH$_1$ and FWH$_2$ from below, and between FWH$_2$ and SN$_4$.
\end{itemize}

\begin{figure}
\centering
\includegraphics[scale=1]{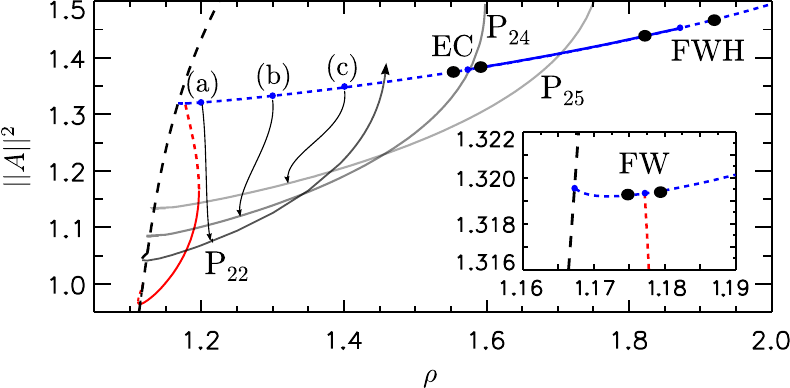}
\caption{(Color online) Bifurcation diagram for $\theta=1.5$. The pattern branch P$_{k_c}$ (red) bifurcates subcritically from HSS at I$^-_{k_c}$, while the branch P$_{2k_c}$ (blue) bifurcates supercritically at I$^-_{2k_c}$. Labels (a)-(c) correspond to the unstable patterns with 32 rolls initially that evolve in time to patterns with different numbers of rolls depending on the value of $\rho$ and lying on new branches of periodic states (gray) labeled by P$_n$, where $n$ is the new roll number. The points where linear stability analysis has been carried out are indicated using the symbol $\bullet$.}
\label{diagrama_EC}
\end{figure}

In Fig.~\ref{diagrama_EC}, we show the bifurcation diagram for $\theta = 1.5$, a value we will use to explore the different instabilities in more detail. For $\theta=1.6$, discussed in Section ~\ref{sec:5}, the results are similar except that P$_{2k_c}$ bifurcates initially subcritically. The temporal evolution indicated by arrows in the figure results from phase slips, as discussed next, and is obtained on a periodic domain of length $L=2\pi n/k_c$, with $n=16$.

\subsection{Eckhaus instability}

For values of $\theta$ and $\rho$ in region III$_{\rm A}$ (see Fig.~\ref{parameter2}), patterns are unstable against long-wavelength perturbations ($q\sim0$), and for this reason the Eckhaus instability is also known as a long-wavelength (LW) instability \cite{Cross,Walgraef_book}. Furthermore, this instability is triggered by a phase instability \cite{Walgraef_book}. For small values of $q$, the least stable branch of eigenvalues $\lambda_1(q)$ has a parabolic shape centered at $q=0$, namely Re$[\lambda_1(q)]\propto|q|^2$, and the instability takes place when the convexity of this eigenvalue branch changes sign. 
\begin{figure}[!h]
\centering
\includegraphics[scale=1]{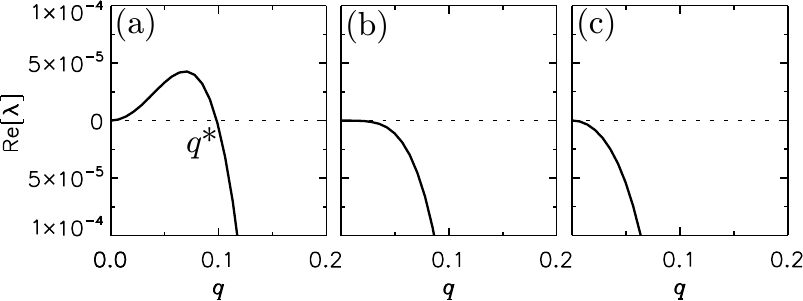}
\caption{The eigenspectrum in the vicinity of the EC instability of the P$_{2k_c}$ branch when $\theta=1.5$, showing Re$[\lambda_1(q)]$ for different values of $\rho$: (a) $\rho=1.57$, (b) $\rho=\rho_{\rm EC}=1.58$, and (c) $\rho=1.59$.}
\label{EC_insta}
\end{figure}

The result of the stability analysis of P$_{2k_c}$ for $\theta=1.5$ and increasing values of $\rho$ as one crosses the 
EC instability threshold is summarized in Fig.~\ref{EC_insta}. In panel (c) $\rho=1.59$ and Re$[\lambda_1(q)]$ is negative for all nonzero $q$. Therefore, P$_{2k_c}$ is stable no matter the wavelength of the perturbation. This situation corresponds to region III$_{\rm B}$ in Fig.~\ref{parameter2}. In panel (b) $\rho=1.58$ and the eigenspectrum flattens around Re$[\lambda_1(q)]=0$, indicating the onset of the EC instability.  Finally in panel (a) $\rho=1.57$ and the eigenspectrum has changed its convexity, indicating that the pattern is now unstable to perturbations with $q\in[0,q^*]$. This property characterizes region III$_{\rm A}$ which extends from EC down to I$^-_{2k_c}$ as $\rho$ decreases.

\begin{figure}[!t]
\centering
\includegraphics[scale=1]{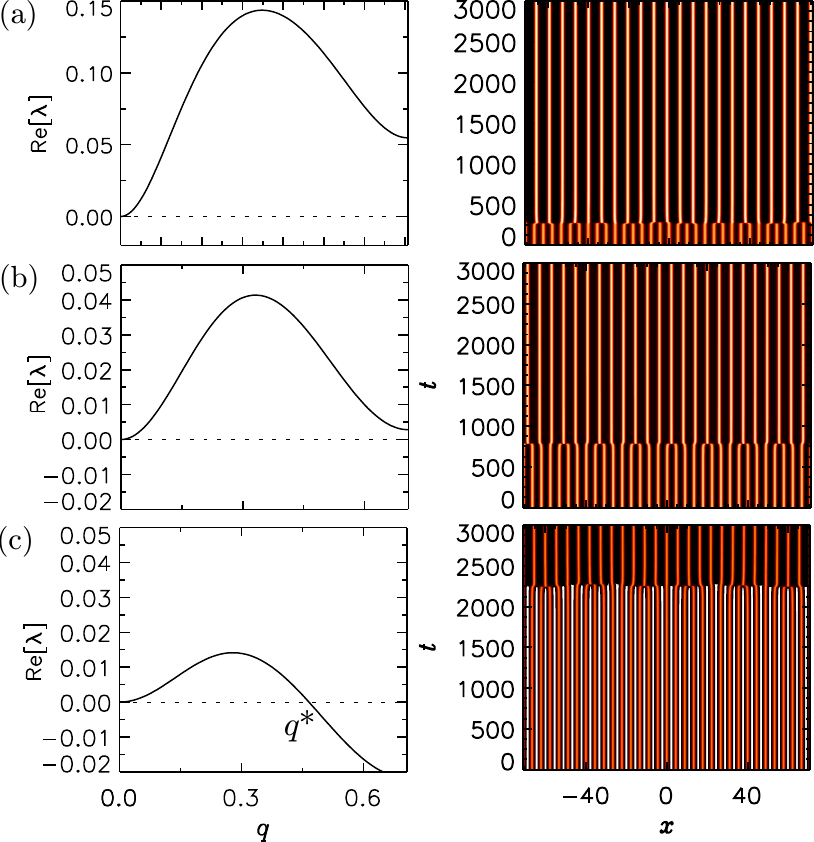}
\caption{Re$[\lambda_1(q)]$ at $\theta=1.5$ in the region of Eckhaus instability and the associated temporal evolution of an unstable initial pattern to patterns of different wavelengths. These new states are shown in gray in Fig.~\ref{diagrama_EC}: an unstable pattern with initially 32 rolls evolves to P$_{25}$ in panel (c) for $\rho=1.4$, to P$_{24}$ in panel (b) for $\rho=1.3$, and to P$_{22}$ in panel (a) for $\rho=1.2$. The left panels show the unstable modes $0<q<q^*$ while the right panels describe the resulting evolution in space-time plots.}
\label{Evol_EC}
\end{figure}

In Fig.~\ref{Evol_EC} the right panels show the temporal evolution of an unstable initial condition along the branch P$_{2k_c}$ together with the real part of the leading eigenvalue $\lambda_1(q)$ [left panels] for different values of $\rho$ in region III$_{\rm A}$. The labels (a)-(c) correspond to different points along the branch P$_{2k_c}$ identified in Fig.~\ref{diagrama_EC}.

For $\rho=1.4$ [Fig.~\ref{Evol_EC}(c)], P$_{2k_c}$ is unstable to perturbations with $q$ in between 0 and $q^*$, and the most unstable mode is that corresponding to maximum growth rate. Time simulations show that after an initial transient during which the pattern appears stable, the wavelength of the pattern suddenly increases to the wavelength of the most unstable mode. The pattern, which initially had 32 rolls, becomes a pattern with 25 rolls that we label P$_{25}$. This new pattern can be tracked in $\rho$ and results in the P$_{25}$ solution branch plotted in Fig.~\ref{diagrama_EC}. 

Reducing the value of $\rho$ further, the P$_{2k_c}$ pattern becomes unstable to any $q\in[0,k'/2]$, with $k'=k_c/2$, and the most unstable wave number increases [Fig.~\ref{Evol_EC}(a)-(b)]. The maximum growth rate Re$[\lambda_1(q)]$ also increases so that the time needed to destabilize the pattern decreases with $\rho$. The final patterns that are reached further beyond the EC instability are P$_{24}$ with 24 peaks in case (b), and the pattern P$_{22}$ in case (a). Once tracked in $\rho$, these stationary patterns generate the solution branches shown in Fig.~\ref{diagrama_EC}.

\subsection{Finite-wavelength instability}

We now characterize the finite-wavelength (FW) instability that allows the pattern P$_{k_c}$ to terminate on P$_{2k_c}$. As already mentioned these locations correspond to a spatial 2:1 resonance located along the line FW in Fig.~\ref{parameter2}. However, the theory described in Refs.~\cite{Armbruster,Proctor,Porter} applies only near the codimension-two case in which the two primary bifurcations from HSS to states with wavenumbers $k_c$ and $2k_c$ occur in close succession. This is not the case here, and we therefore employ the numerical technique of the previous section to compute the location of the FW bifurcation when this occurs in the fully nonlinear regime.

\begin{figure}[!t]
\centering
\includegraphics[scale=1]{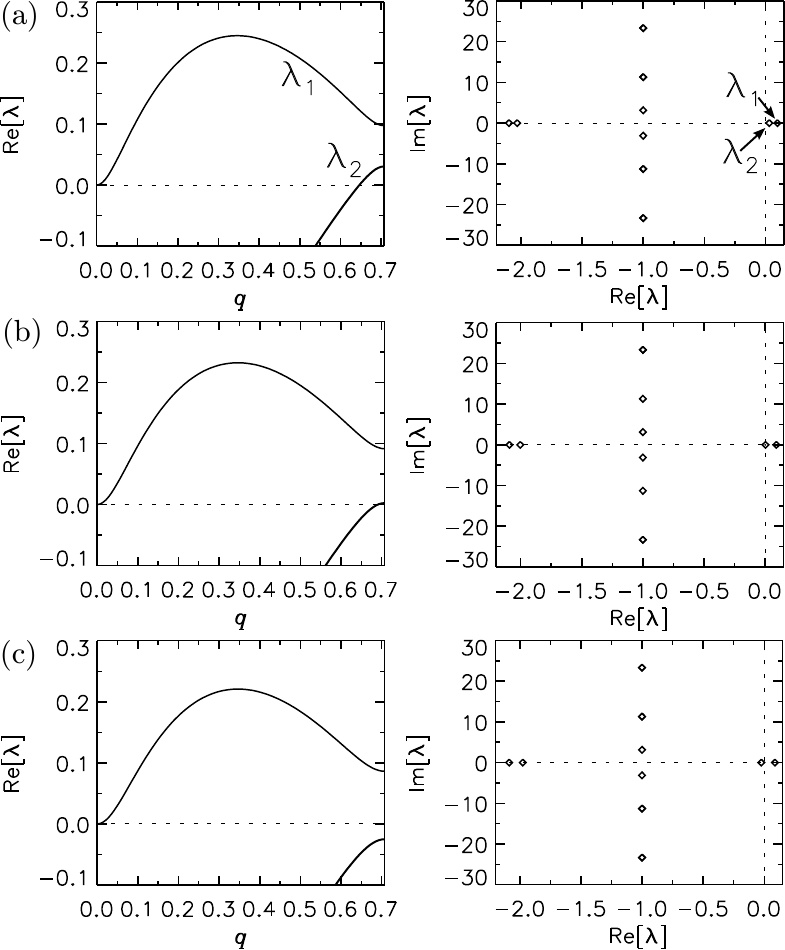}
\caption{The eigenspectrum of P$_{2k_c}$ in the vicinity of the FW instability when $\theta=1.5$, showing the first two branches Re$[\lambda_1(q)]$ and Re$[\lambda_2(q)]$ for different values of $\rho$: (a) $\rho=1.175$, (b) $\rho=\rho_{\rm FW}\approx 1.177$, and (c) $\rho=1.179$.}
\label{instaFW}
\end{figure}

If $k'=2k_c$ is the wavenumber of P$_{2k_c}$, the FW bifurcation is characterized by a branch of eigenvalues $\lambda_2(q)$ having a parabolic shape centered at $q=k'/2$, i.e., Re$[\lambda_2(q)]\propto|q-k'/2|^2$, which crosses Re[$\lambda_2(q)]=0$ at $q=k'/2$. This transition is shown in Fig.~\ref{instaFW} for $\theta=1.5$ and for three values of $\rho$ in the vicinity of the FW bifurcation [see the inset in Fig.~\ref{diagrama_EC}]. The real part of the two leading eigenvalues $\lambda_1(q)$ and $\lambda_2(q)$ is shown in the left panels, while the right columns show the full eigenspectrum at $q=k'/2=k_c$. In any case Re[$\lambda_1(q)$] is positive for all the range $q\in[0,k'/2=k_c]$, and therefore P$_{2k_c}$ is unstable against Bloch modes with $q\in[0,k_c]$, i.e. in this regime P$_{2k_c}$ is EC unstable. The FW transition is triggered by the second eigenvalue $\lambda_2$ centered at $q=k'/2$.  In (a) $\rho<\rho_{\rm FW}$, and a portion of the branch Re[$\lambda_2(q)$] is positive, with its maximum occurring at $q=k'/2$. Therefore, in this case P$_{2k_c}$ is unstable to the most unstable mode, i.e. $q=k'/2=k_c$, and therefore to P$_{k_c}$, in addition to the unstable EC mode. In (b) $\rho=\rho_{\rm FW}$, and the maximum growth rate Re[$\lambda_2(q)$] at $q=k'/2$ vanishes, as can be appreciated by looking at the corresponding eigenspectrum in the right column. This point therefore corresponds to presence of the FW bifurcation. Finally, panel (c) shows the situation at $\rho>\rho_{\rm FW}$, where Re[$\lambda_2(q)$] is negative for all $q$, and the P$_{2k_c}$ pattern is FW stable.

\subsection{Finite-wavelength-Hopf instability}

For values of $\theta$ and $\rho$ in region III$_{\rm C}$ patterns undergo a finite-wavelength-Hopf instability, hereafter FWH. In contrast to the homogeneous Hopf bifurcation which occurs with $q=0$, this Hopf bifurcation sets in with a finite wave number $q\neq0$, here $q=k_c$. In the former case, patterns which are Hopf unstable will oscillate with a uniform amplitude and temporal period $T=2\pi\omega$, with $\omega={\rm Im}(\lambda_2(0))={\rm Im}(\lambda_3(0))$. Here $\lambda_{2,3}(0)$ are the Hopf modes. In the FWH case, however, patterns oscillate both in time and in space, and this is why this instability is also referred to as a {\it wave instability} (WI) \cite{Cross, Walgraef_book,Hildebrand,Epstain1,Epstain2}.

\begin{figure}[!t]
\centering
\includegraphics[scale=1]{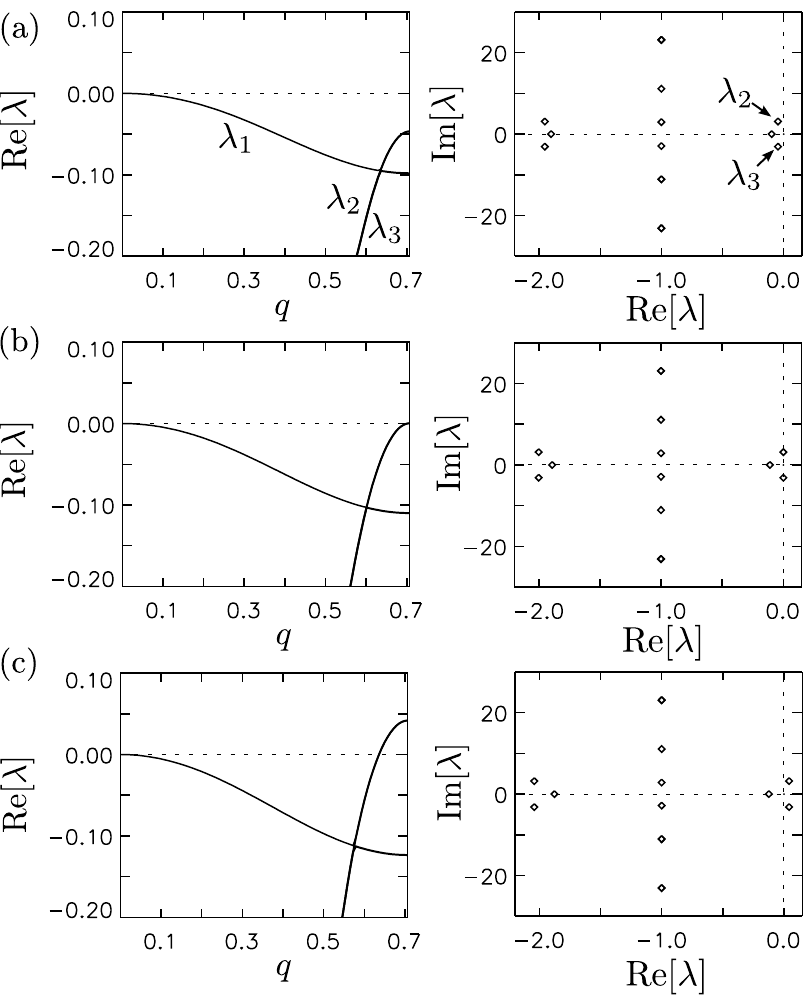}
\caption{Hopf bifurcation of P$_{2k_c}$ at $\theta=1.5$ showing (left panels) Re$[\lambda(q)]$ for different values of $\rho$: (a) $\rho=1.82$, (b) $\rho=\rho_{\rm FWH}=1.87$, and (c) $\rho=1.92$. The right panels show the corresponding eigenspectrum at $q=k_c$, the onset wave number.}
\label{Hopf_insta}
\end{figure}

\begin{figure}[!t]
\centering
\includegraphics[scale=1]{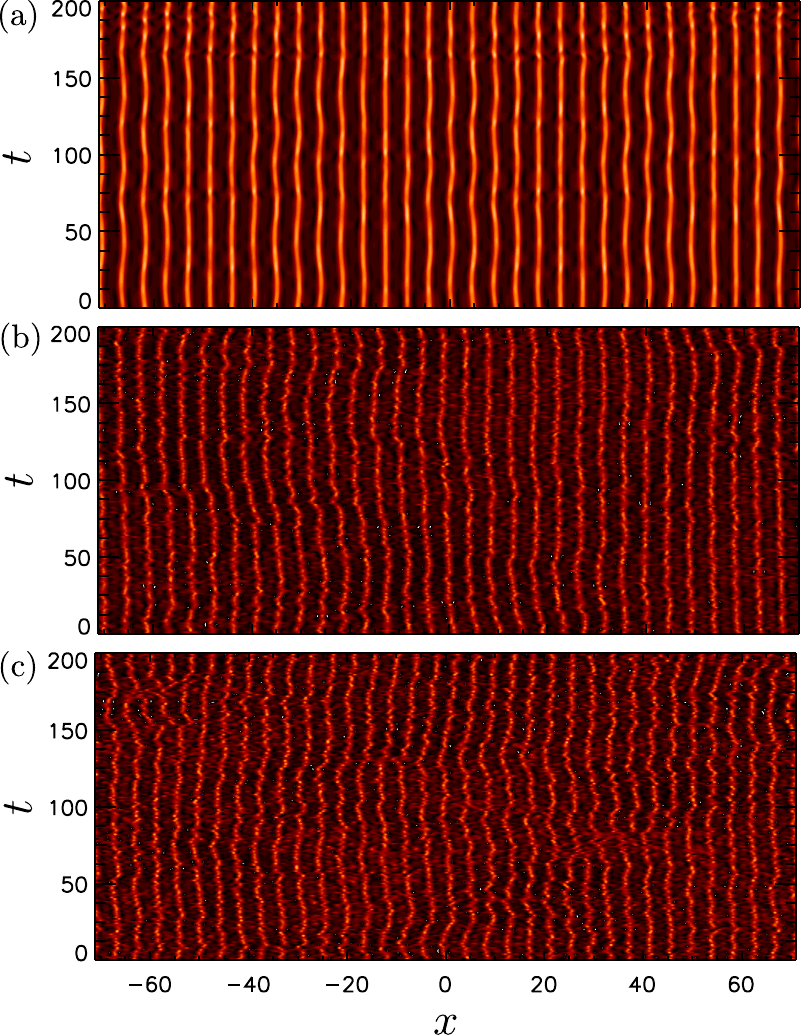}
\caption{Time evolution of the oscillating patterns for $\theta=1.5$ and (a) $\rho=1.9$, (b) $\rho=2.1$ and (c) $\rho=2.3$.}
\label{Hopf_evol}
\end{figure}

In Fig.~\ref{Hopf_insta} the real part of the three leading eigenvalues (left) and the full eigenspectrum at $q=k'/2=k_c$ (right) are plotted when crossing the FWH bifurcation at $\theta=1.5$ [see Figs.~\ref{parameter2} and \ref{diagrama_EC}]. In panel (a) $\rho=1.82$, and the real parts of $\lambda_2(q)$ and $\lambda_3(q)$ are both negative, with a parabolic shape centered at $q=k'/2=k_c$. In fact these eigenvalues are complex conjugates of one another, as can be seen in the full eigenspectrum for $q=k_c$ shown in the right panel. This is the situation in region III$_{\rm B}$ where P$_{2k_c}$ is FWH stable. In panel (b) $\rho=\rho_{\rm FWH}=1.87$ and the real part of the complex conjugate eigenvalues $\lambda_{2,3}(q)$ vanishes at $q=k_c$, indicating the onset of the FWH bifurcation. Finally, in (c) $\rho=1.92$, and the real part of the eigenvalues is now positive and P$_{2k_c}$ starts to oscillate, not only in time but also in space. This is the situation of region III$_{\rm C}$ shown in Fig.~\ref{parameter2}. 

In Fig.~\ref{Hopf_evol}, we show the resulting oscillatory states for different values of $\rho$ in region III$_{\rm C}$ when $\theta=1.5$. For $\rho=1.9$ [see panel (i)], the amplitude of P$_{2k_c}$ oscillates non-uniformly not only in time but also in space resulting in zig-zag motion whose amplitude grows with increasing $\rho$ as seen in panel (b). Finally, in panel (c), for $\rho=2.4$, the pattern exhibits much complex dynamics including phase slips at which peaks merge or splitresulting in fluctuations in the total number $n(t)$ of rolls in the domain at any one time. A complete description and understanding of the dynamics of these oscillatory states in time and space involves interaction with the marginally stable $q=0$ mode [Fig.~\ref{Hopf_insta} and \cite{CoxMatthews}] and is beyond the scope of this paper.

\section{Conclusions}
\label{sec:6}
In this paper we have studied the bifurcation structure and stability properies of spatially periodic patterns arising in the LL model in the anomalous GVD dispersion regime.

Linear stability theory predicts that the HSS solution becomes modulationally unstable at $I_0=I_c=1$ to a pattern with a critical wave number $k_c=\sqrt{2-\theta}$, namely P$_{k_c}$ \cite{lugiato_spatial_1987,Tlidi_96}. A weakly nonlinear analysis has allowed us to obtain a perturbative description of this pattern in the neighborhood of this bifurcation. From this calculation one finds that P$_{k_c}$ emerges supercritically for $\theta<41/30$ and subcritically when $\theta>41/30$, where $\theta=41/30$ corresponds to a degenerate HH point. 

This analytical approximation for the pattern P$_{k_c}$ around the MI point (or equivalently: HH) has been used as an initial condition in a numerical continuation algorithm that allowed us to track the pattern solutions to parameter values away from the bifurcation point. Using this method, we have studied the bifurcation structure of spatially periodic patterns as a function of $\rho$ for different values of the detuning $\theta$. In doing so, we have found that for low $\theta$ patterns arising from the MI bifurcation reconnect with the HSS for larger values of the pump intensity $I_0$, at I$^{+}_{k_c}$. In addition, harmonic patterns with wave numbers $nk_c$, $n=2,4,\dots$ also bifurcate from the HSS, P$_{2k_c}$ at I$^{\pm}_{2k_c}$, P$_{4k_c}$ at I$^{\pm}_{4k_c}$, etc. With increasing $\theta$ these these two types of patterns connect pairwise in a 2:1 spatial resonance, for example P$_{k_c}$ with P$_{2k_c}$ and P$_{2k_c}$ with P$_{4k_c}$. We have referred to these bifurcation points as finite-wavelength (FW) instabilities, and computed their location via numerical Floquet analysis. This FW bifurcation originates in the codimension-two point X, which appears to organize these connections. Finally, as $\theta \rightarrow 2$ and $k_c \rightarrow 0$ the bifurcation structure of patterns transforms into foliated snaking of localized structures \cite{Parra_Rivas_P1}, as a pattern with infinite wavelength corresponds in effect to a single peak localized structure in a finite size system.

We have provided an almost complete discussion of the various possible secondary bifurcations in the parameter space $(\theta,\rho)$ of the LL equation, mapping the different dynamical regions for the patterns P$_{k_c}$ and P$_{2k_c}$. In particular, patterns corresponding to P$_{2k_c}$ were found to undergo Eckhaus and finite-wavelength-Hopf instabilities, in addition to the FW instability, and these were found to lead to rich and complex dynamics. Several significant but higher codimension bifurcation were also identified, but a detailed study of these remains for future work.

While we have focused our study on patterns with the critical wave number $k_c$ determined by the onset of the MI, and its harmonics, we have confirmed that similar behavior also occurs for patterns with wave number $k\ne k_c$ that also emerge from the HSS solution when $I_0>I_c$. Together with the instabilities described in this work, other bifurcations such as an FW with $q=k/3$ are also known to exist \cite{Perinet_Eckhaus}. A detailed study of secondary instabilities of patterns with arbitrary wave number $k$ are beyond the scope of this paper, however, and are likewise left to future work.\\

\acknowledgments      
We acknowledge support from the Research Foundation--Flanders (FWO-Vlaanderen) (PPR), internal Funds from KU Leuven (PPR), the Belgian Science Policy Office (BelSPO) under Grant IAP 7-35, the Research Council of the Vrije Universiteit Brussel, and the Agencia Estatal de Investigaci\'on (AEI, Spain) and Fondo Europeo de Desarrollo Regional under Project ESoTECoS, Grants No. FIS2015-63628-C2-1-R (AEI/FEDER,UE) (DG) as well as the National Science Foundation under grant DMS-1613132 (EK).

\end{document}